\documentclass[12pt]{iopart}

\usepackage{graphicx}
\newcommand \beq{\begin{eqnarray}}
\newcommand \eeq{\end{eqnarray}}
\begin{document}

\title[Bulk and Spectral Observables in Lattice QCD]
{Bulk and Spectral Observables in Lattice QCD}

\author{T Hatsuda}

\address{Department of Physics, The University of Tokyo, Tokyo 113-0033, Japan}
\eads{{hatsuda@phys.s.u-tokyo.ac.jp}}
\begin{abstract}
We review recent developments in lattice siumulations of the equation of state, 
order of the thermal phase transition and the determination of the pseudo-critical temperature in (2+1)-flavor QCD. Owing to the increasing computer power, 
new argothithms, and improved fermion formulations,  studies of bulk QCD matter are
 approaching to the stage of precision science. We also review recent lattice studies on 
 the spectral properties of heavy quarkoniums inside the quark-gluon plasma (QGP). 
 Although they are still in an exploratory stage, interesting physics in relation to 
 the  strongly correlated QGP is being extracted. 
  \end{abstract}


\section{Introduction}

\begin{figure}[b]
\begin{center}
\includegraphics[width=8cm]{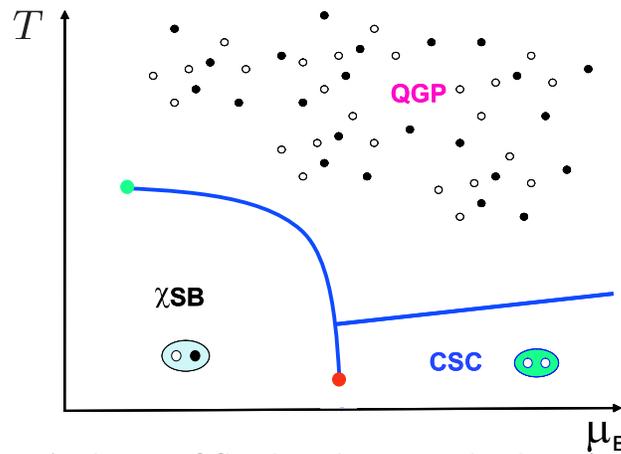}
\caption{\label{T-mu}
A schematic QCD phase diagram in the plane of 
temperature $T$ and baryon chemical potential $\mu_{_{\rm B}}$.
 Solid lines indicate the first order phase boundaries.
 Several critical points at which the first order lines
  terminate may exit, e.g. the high $T$ critical point \cite{asakawa-yazaki}
     and high $\mu_{_{\rm B}}$ critical point \cite{yamamoto}.}
\end{center}
\end{figure}

 One of the main goals of the lattice QCD studies
  is to make first principle analysis of hot/dense QCD
  and to supply reliable inputs
     to quark-gluon plasma (QGP) phenomenologies \cite{QGP}.
  Shown in Fig.\ref{T-mu}  is a schematic
  QCD phase diagram indicating  three basic
  phases; the chiral symmetry broken ($\chi$SB) phase,
   the color superconducting (CSC) phase and
    the QGP phase.
   Precise locations of the phase boundaries and the 
   critical points as well as the dynamics 
  in each phase should be determined by
  non-perturbative method such as the lattice QCD.

  QCD partition function  in a finite box with a spatial volume $V$
  and the  lattice spacing $a$ can be written as
  \begin{eqnarray}
 Z(T, \mu; a,V)
 = {\rm Tr} \left[ {\rm e}^{-(\hat{H}-\mu \hat{N})/T} \right] 
 = \int [dU]\ F(U)\ {\rm e}^{-S_g(U)} , 
 \label{eq:Z}
\end{eqnarray}
  where $U$ is the gauge field defined as an element of 
  $SU(3)_{\rm c}$ and $F(U)$ denotes the quark contribution to $Z$.
  Eventually, we need
    to take the continuum limit ($a \rightarrow 0$) and
     the  thermodynamic limit ($V \rightarrow \infty)$ to 
    extract physical observables in the real world.

\begin{figure}[b]
\begin{center}
 \includegraphics[width=8.5cm]{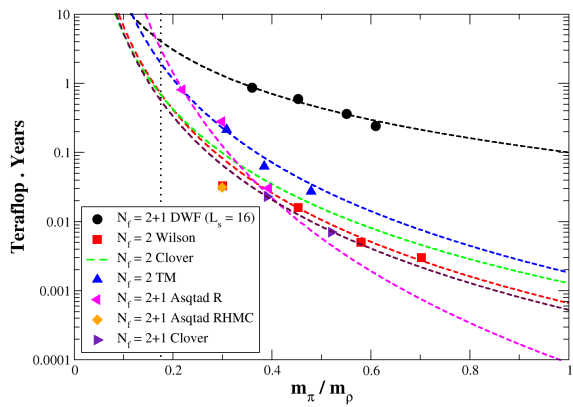}
 \caption{\label{HMC}
Updated Berlin wall plot which shows  the cost to generate
 1000 independent gauge configurations on
 $24^3 \times 40$ lattice with $a=0.08$ fm in various fermion
 formulations with modern algorithms \cite{clark06}.
  }
\end{center}
\end{figure}

   In the past few years, there have been considerable
 progress in lattice QCD approach: (i) available computer speed 
  becomes as fast as 50 Tflops, (ii)
    improved fermion formulations
   have been tested extensively  (such as staut, asqtad and p4 improved actions 
   for staggered quarks and the clover action for Wilson quark),
  (iii)  new fermion formulations with good chiral properties are
   started to be used (such as domain wall and overlap fermions),
  and  (iv) new algorithms for full QCD simulations are proposed and 
   implemented  (such as the rational hybrid Monte Carlo method (RHMC)
  and the domain-decomposition hybrid Monte Carlo method (DDHMC)
  \cite{fermion-alg}).
   Shown in Fig.\ref{HMC} is so called the
    Berlin wall plot where the computational cost in 
     unit of Tflops$\cdot$year is plotted against 
    $m_{\pi}/m_{\rho}$.
  The quark mass dependence of the cost becomes weaker than
   before, and simulations with
   realistic quark masses may be done in a reasonable
   amount of time by using O(50) Tflops machines.   

  In the following, we will focus on the 
  bulk and spectral properties associated with the 
  QCD phase transition at finite $T$ with $\mu_{_{\rm B}}=0$.
   For the developments in lattice QCD at
  finite chemical potential, see the recent review \cite{schmidt06}.
 
\section{Equation of state in (2+1)-flavor QCD}

\begin{figure}[b]
\begin{center}
\includegraphics[width=9cm]{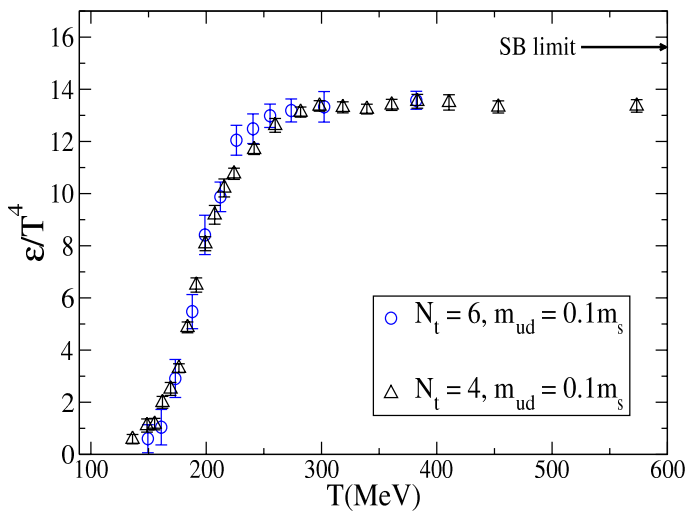}
\caption{\label{milc}
 Energy density $\varepsilon$ as a function of $T$ 
 for (2+1)-flavor QCD with asqtad improved action
 on the lattice, 
  $N_t \times N_s^3 = 4 \times (12^3-16^3)$ and $6 \times (12^3-20^3)$ 
  \cite{milc_eos_06}. 
 The strange quark mass $m_{\rm s}$ is about the physical value,
  while the light quark mass $m_{\rm ud} \simeq 0.1 m_s$
  corresponding to $m_{\pi}/m_{\rho} \simeq 0.3$. 
     The Sommer scale $r_1=0.138 (7)(4)$ fm is used to set the scale.
   Lattice spacing at  $T\simeq 200$ MeV is 0.247 (0.165) fm for $N_t=4\ (6)$.}
\end{center}
\end{figure}

The equation of state at finite $T$ characterizes the bulk QCD matter.
 Also it becomes an fundamental input together with the 
  transport coefficients \cite{sakai06}
 to the relativistic hydrodynamics for  QCD fluid \cite{hirano06}.
 Shown in Fig.\ref{milc} is the 
 state-fo-the-art calculation of the 
  energy density $\varepsilon$ as a function of $T$
 in (2+1)-flavor QCD  with asktad improved staggered quarks. 
 (See also the recent results for $3$-flavor QCD with the p4 action
 \cite{karsch_eos_06} and for $(2+1)$-flavor QCD with the stout action
 \cite{aoki_eos_06}.)

 The figure shows that $\varepsilon/T^4$  
 increases rapidly at $T \sim 200$ MeV and approaches 
  the Stefan-Boltzmann (SB) limit from below.
   The deviation from the SB limit at high $T$
  is due to the quark-gluon interaction
   which decreases only logarithmically as $T$ increases.
  Quantitative understanding of 
  the deviation is a non-trivial issue and cannot be treated
   in naive thermal perturbation \cite{rebhan04}.

\section{Order of the transition in (2+1)-flavor QCD}

 Using the pressure 
  $P(\vec{K})= (T/V) \cdot {\rm ln} Z$ and  
  a vector 
   $\vec{K}\equiv (T, \mu, m_q, a, V, \cdots )$,
   one may define the $n$-th order transition
 as the one where non-analyticity  of $P$ in the 
  thermodynamic limit ($V \rightarrow \infty $)
  first appears in the $n$-th derivative of $P$ with respect
   to $\vec{K}$. For example, if 
   $(\partial/\partial T)P=s$ is discontinuous at certain
    $T$, it is the first order transition.
  On the other hand, if
  $(\partial/\partial T)P$ is continuous and 
   $(\partial/\partial T)^2 P=(\partial/\partial T) s=c_{_V}/T$
    diverges at certain $T$, it is an example of the 
    second order transition.
  If smooth phase change takes place without 
  the non-analyticity of $P$, it is the crossover.
  To make a distinction between the phase transition and the crossover
 from the numerical simulations, 
 the scaling behavior of certain observables as a function of 
 $V$  needs to be  studied (the finite size scaling analysis) 
 \cite{ukawa93}.  In the (2+1)-flavor QCD, the relevant
  observable is the chiral susceptibility
  whose volume dependence at its maximum reads 
\begin{eqnarray}
 \chi_m  = (\partial/\partial m_{\rm ud})^2 P 
 \ \ \rightarrow \ \  V^{\alpha} \ \ {\rm at\ the\ peak}, 
\end{eqnarray}
where $\alpha \simeq 1\ (2/3)$ for the first (second) order transition,
 and  $\alpha \simeq 0$ for crossover.

\begin{figure}[b]
\begin{center}
\includegraphics[width=6.5cm]{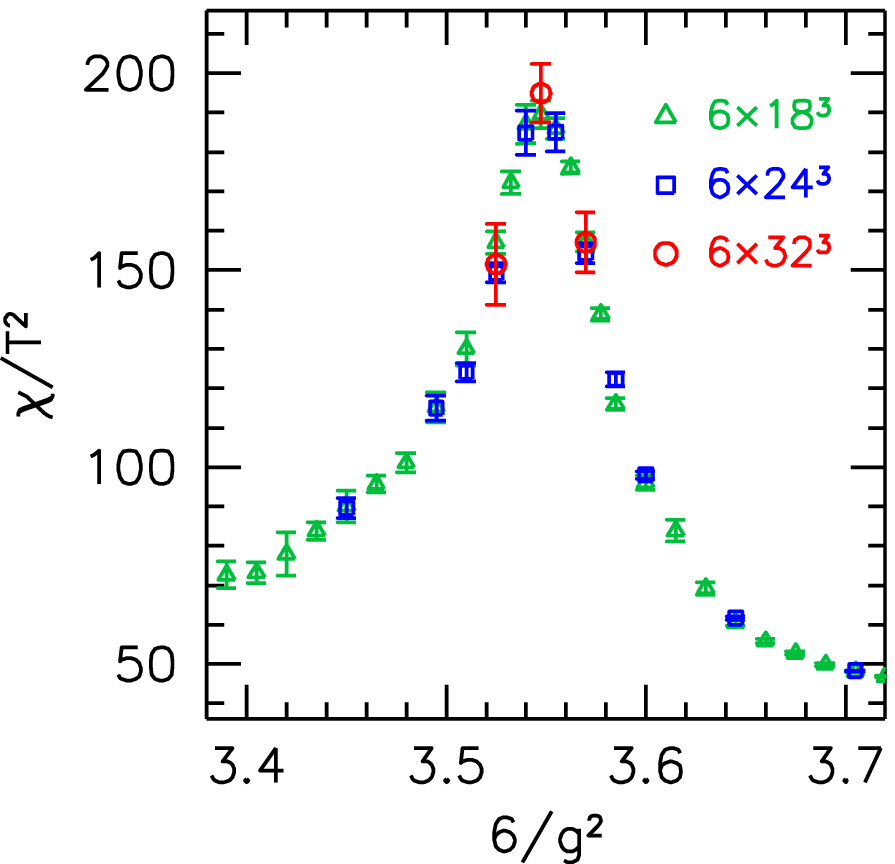}
\hspace{0.8cm}
\includegraphics[width=7.5cm]{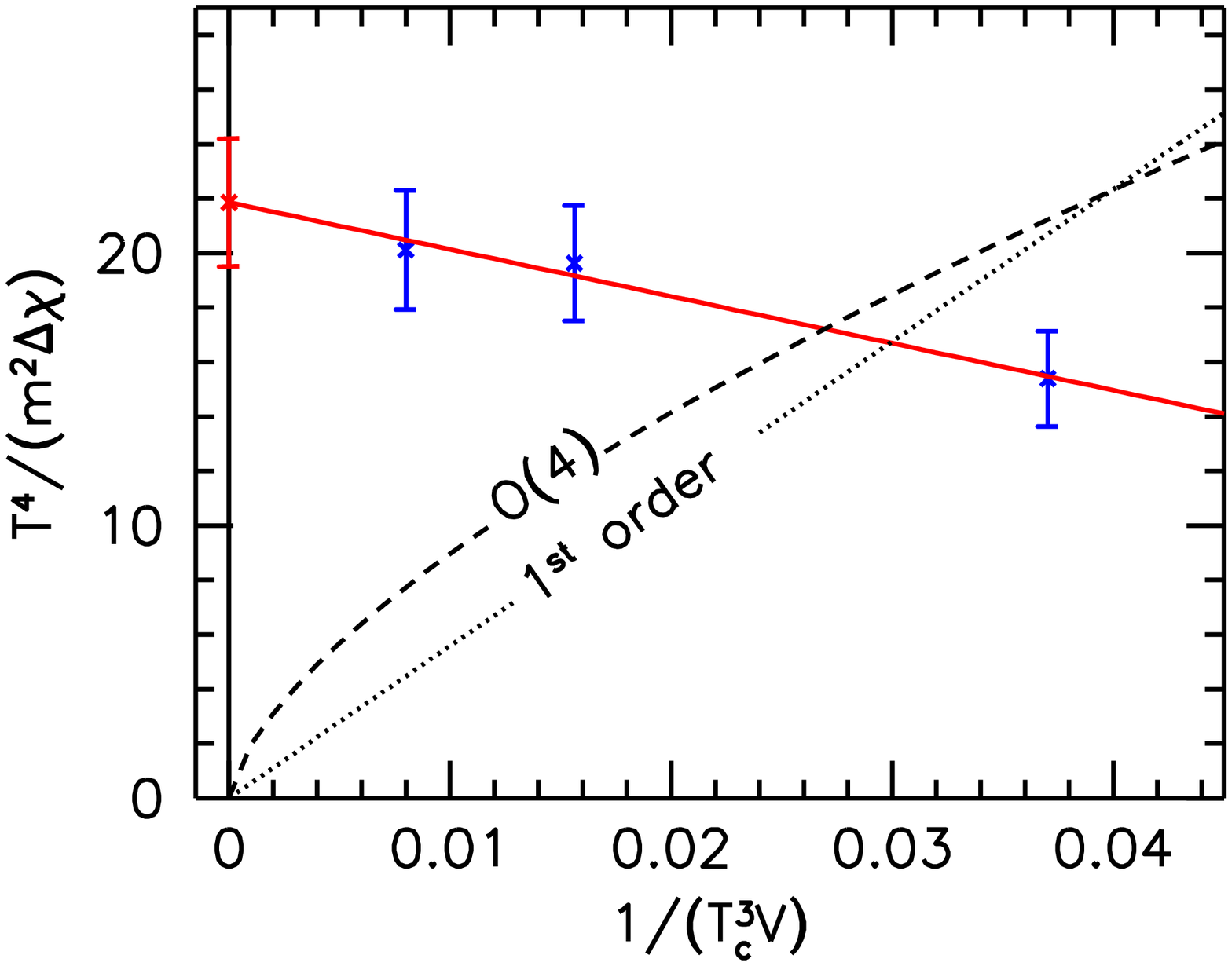}
\caption{Left: The chiral susceptibility in (2+1)-flavor QCD
 as a function of the bare coupling, $6/g^2$, for
 different values of the spatial volume, 
 $N_t^3=V/a^3 = 18^3, 24^3, 32^3$.
 The stout improved staggered fermion action is employed and the
 quark masses are chosen to reproduce
 $m_{_K}/m_{\pi}$=3.689 and $f_{_K}/m_{\pi}$=1.185.
Right: The volume dependence of the peak of the 
subtracted chiral susceptibility in the continuum limit.
 Dotted (dashed) line is an expectation
  from the first order (second order) phase transition.
 Figures are taken from \cite{aoki_sus_06}.}
\label{aoki_sus}
\end{center}
\end{figure}

 Shown in the left panel of Fig.\ref{aoki_sus} is the 
 dimensionless chiral susceptibility $\chi_m/T^2$ 
 around its peak for different values of the 
  dimensionless volume $N_s=V/a^3 = 18^3, 24^3, 32^3$
  for (2+1)-flavor QCD with realistic masses of u, d, s quarks \cite{aoki_sus_06}.
 There is no appreciable volume dependence for the height
  of the susceptibility, which suggests the 
  crossover nature.
  The right panel of Fig.\ref{aoki_sus} shows
  the subtracted chiral susceptibility in the 
   continuum limit  $\Delta \chi$ ( = the peak of  $\chi_m(T)-\chi_m(0)$)
  in a renormalization group  invariant combination 
  $T^4/(m_{\rm ud}^2 \Delta \chi)$ as a function of the 
   dimensionless inverse-volume, $1/(T_{\rm c}^3 V)$.  
  The height of the susceptibility approaches to a non-vanishing constant
   in the thermodynamic limit and has different 
   scaling behavior from those of first or second order
    transitions. This gives a strong evidence
     that the hadron-QGP transition at finite $T$ with zero chemical potential
    is a crossover in the real world.

\section{Pseudo-critical temperature for (2+1)-flavor QCD}

\begin{figure}[b]
\begin{center}
\includegraphics[width=12cm]{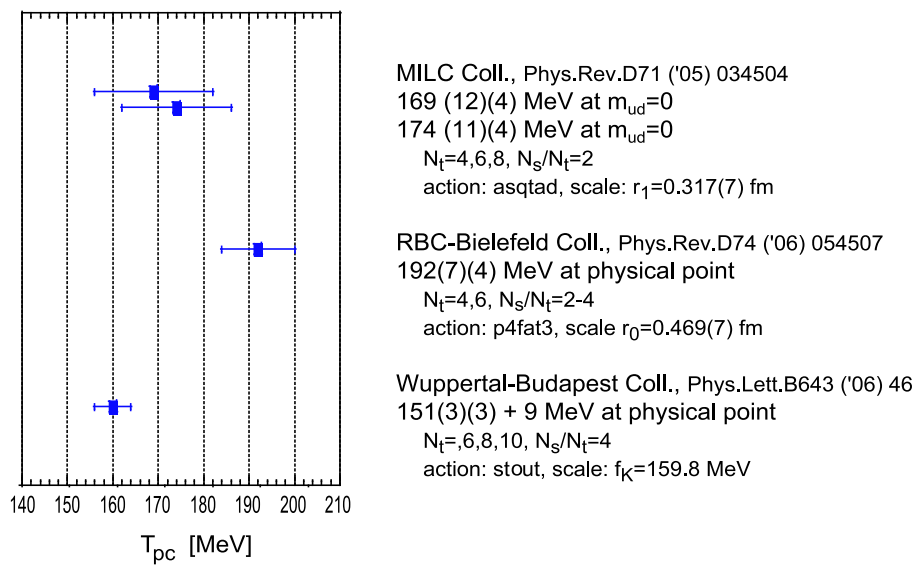}
\caption{\label{T_pc}
Recent determinations of the (pseudo-)critical temperature
for (2+1)-flavor QCD with  staggered quarks.
 As for the lowerst bar, we add 9 MeV  to
 make a comparison to other results
 obtained from $\chi_m/T^2$ \cite{WB06_tc}.
 }
\end{center}
\end{figure}

 Since the hadron-QGP transition in the real world is 
 crossover,
 the ``critical" temperature $T_{\rm c}$ is not 
 a well-defined concept.
 Nevertheless, one may introduce a ``pseudo-critical"
  temperature $T_{\rm pc}$ as a peak position of certain
  susceptibility. The actual value of
  $T_{\rm pc}$ is different for different choice of
   susceptibility, e.g. 
   $\chi_m=({\partial}/{\partial m_{\rm ud}})^2 P$,
 $\chi_{T}=({\partial}/{\partial T})^2 P$, and 
 $\chi_{mT}=({\partial^2}/{\partial m \partial T}) P$.
  Note also that multiplying an arbitrary function of $m_{\rm ud}$
  and $T$ to the susceptibilities can change the value of
  $T_{\rm pc}$.  
 
 Shown in Fig.\ref{T_pc} is a summary of
  (pseudo-)critical temperature recently calculated 
   from the dimensionless chiral
 susceptibility, $\chi_m/T^2$, for (2+1)-flavor QCD.
 Extrapolation to the continuum limit is taken by 
 using the data for several values of $N_t$.
 The upper two bars are $T_{\rm c}$  
   with different extrapolations to the chiral limit
   $m_{\rm ud}=0$ by MILC collaboration  \cite{MILC06_tc}. 
 The middle bar is $T_{\rm pc}$ extrapolated to the 
  realistic  $m_{\rm ud}$ by RBC-Bielefeld collaboration \cite{RBCB06_tc}.
 The lower bar is $T_{\rm pc}$ calculated at the 
  realistic $m_{\rm ud}$ by Wuppertal-Budapest collaboration \cite{WB06_tc}.
 Note that different improved actions 
   and different scale determinations are adopted in three groups.
 When extracting  $T_{\rm pc}$ 
  in MeV unit, we need to use the chain rule:
 \begin{eqnarray}
T_{\rm pc} = \left( \frac{T_{\rm pc}(a)}{a^{-1}} \right)_{a \rightarrow 0}
\cdot
 \left( \frac{a^{-1}}{M(a)} \right)_{a \rightarrow 0} 
\cdot
 M_{\rm ``exp"}.
 \end{eqnarray}
 Here $M$, which  has a dimension of mass,
  is chosen either to be a direct physical 
 observable (such as $m_{\rho}$, $f_\pi$ and 
 $f_{_K}$)  or to be a semi-empirical quantity
 (such as the Sommer scales $1/r_0$ 
 and $1/r_1$ defined from the heavy-quark potential
  through  $\left. r^2 \ dV(r)/dr \right|_{r=r_0} = 1.65$
   and $\left. r^2 \ dV(r)/dr \right|_{r=r_1} = 1.0$). 

 There are a number of issues to be studied further
  before drawing definite conclusion on the value of 
  $T_{\rm pc}$ in the real world.
   In particular,
  (i)  careful continuum extrapolation ($a\rightarrow 0$) 
   by the data with larger $N_t$,
   (ii) determination of $T_{\rm pc}$ in the case of 
    Wilson fermion \cite{maezawa06},
   and (iii) resolving the difference between the Sommer scale
    in the continuum limit
   obtained from $a$ being determined by the 2S-1S mass splitting of the 
    bottomonium with staggered quark ($r_0=0.468(7)$ fm)
    \cite{HPQCD-UKQCD} and that
     obtained from $a$ being determined by the kaon mass with Wilson quark
      ($r_0=0.516(21)$ fm) \cite{CP-PACS-JLQCD}.

\section{Heavy quarks inside QGP}

 Heavy quarks in the real world (such as the 
 charm and bottom) have finite masses and 
 may receive substantial kick from the light 
  plasma-constituents which have typical
  energy of about $3T$.  
  Therefore, it is important to examine the 
   dynamical correlations (both spatial and 
   temporal) of the heavy quarks inside QGP.
   It is also relevant to study the modification of the 
   heavy quarkonium properties in hot QCD matter \cite{MS}.

 \begin{figure}[b]
\begin{center}
\includegraphics[width=8.5cm]{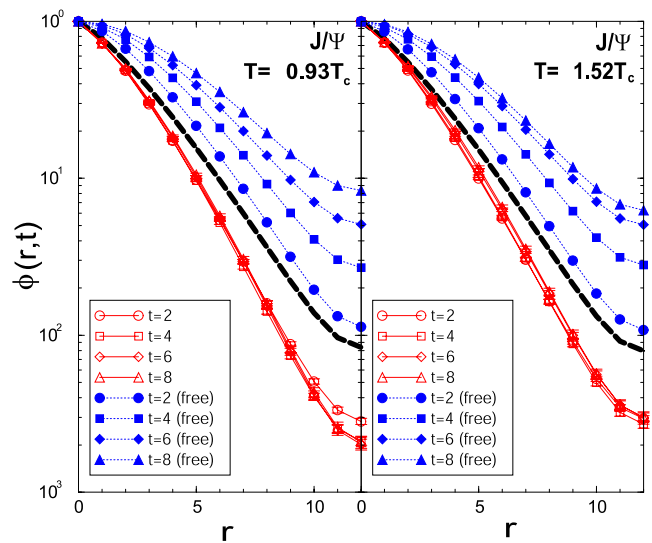}
\caption{\label{umeda}The equal-time Bethe-Salpeter wave function normalized
 at the origin as a function of the spatial separation $r$ between
 the charm  and the anti-charm. Blue (red) points correspond
  to free (interacting) quarks \cite{umeda}. 
  The black dashed line is the wave function initially prepared 
  at $\tau=0$. Note that $t \equiv \tau$ in this figure.
   }
\end{center}
\end{figure}

\subsection{Charmonium wave function in quenched QCD}

 An attempt to study the $J/\Psi$ wave function
 at finite temperature in quenched QCD ($F(U)$ is take to be unity
 in Eq.(\ref{eq:Z}))
 on an anisotropic lattice was performed
 by Umeda et al. \cite{umeda}.
 They have studied the equal-time Bethe-Salpeter
  wave function ${w}({\vec r},\tau) $ at finite $T$ in the Coulomb gauge.
 In Fig.\ref{umeda}, the normalized wave function,
 $\phi({\vec r},\tau)={w}_i({\vec r},\tau)/{w}_i({\vec 0},\tau)$
  is plotted below and above $T_c$ for various time slices.
  The black dashed line is the wave function prepared 
  at $\tau=0$.
  For free quarks without gauge interactions, the wave
   function becomes broader as the imaginary time $\tau$ increases.
   On the other hand, the wave function 
   for the interacting system is almost unchanged even at $T=1.52 T_c$,
    which suggests that
    $J/\Psi$ may survive as a bound state in the deconfined phase.

\subsection{Charmonium spectral function in quenched QCD}

The spectral functions of hadronic correlators
 give us another information on 
 hadronic modes at finite  $T$ as originally suggested in 
  \cite{HK85}.
 The spectral function $\sigma (\omega, {\vec p})$ for 
  hadronic correlation may be defined through the 
  spectral decomposition:
 \begin{eqnarray}
 \label{eq:7.mixed-sum}
 D ( \tau , {\vec p} ) 
 = \int_{-\infty}^{+\infty} {{\rm e}^{-\tau \omega} \over
 1 \mp {\rm e}^{-\omega/T}} \ \sigma (\omega, {\vec p})
 \ d\omega \ \ \ (0  \le   \tau < T^{-1}).  
\end{eqnarray}   
  The maximum entropy method (MEM)
   provides an efficient and  powerful   
   way to obtain a unique $\sigma$ from the lattice data $D$
   without making  a priori parameterization \cite{AHN01}.

\begin{figure}[b]
\begin{center}
\includegraphics[width=7.5cm]{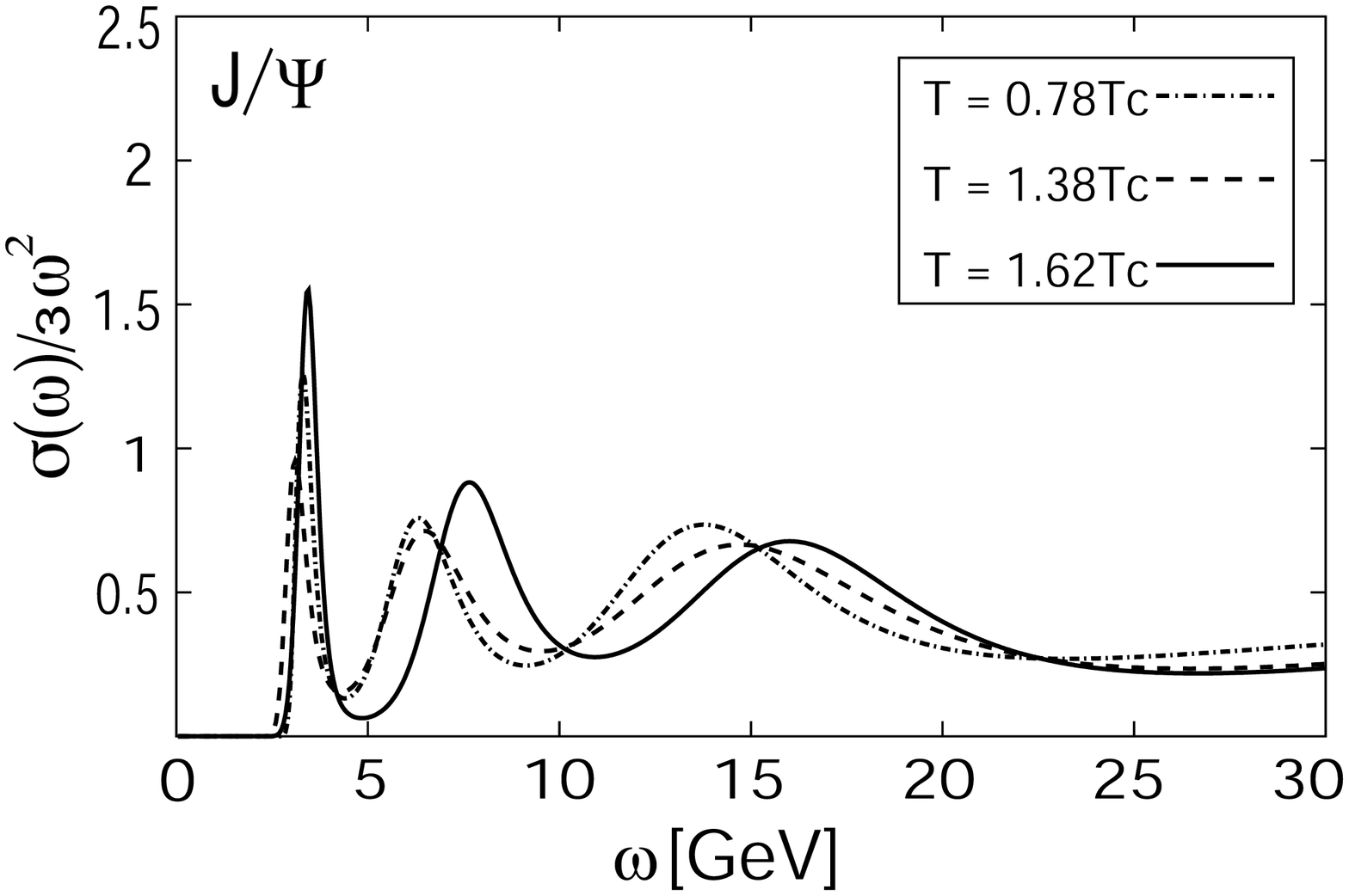}
\hspace{1cm}
\includegraphics[width=6.8cm]{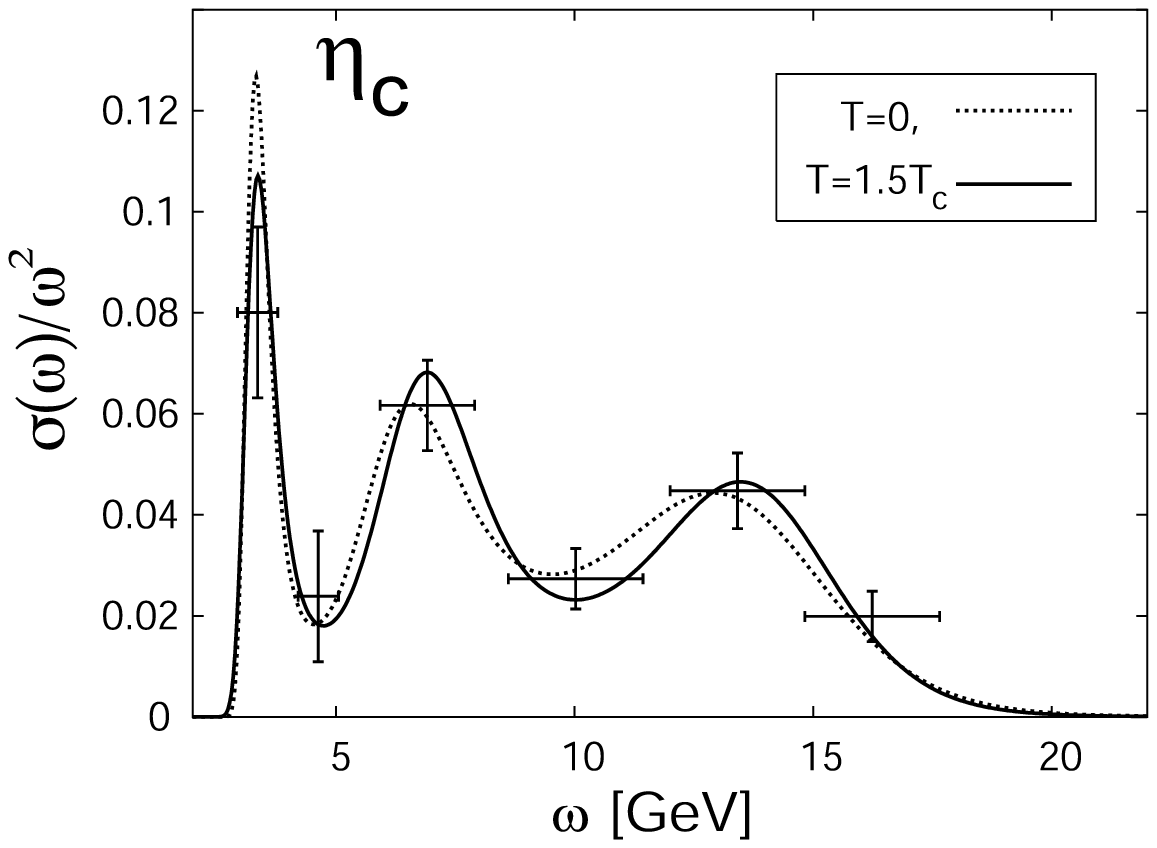}
\caption{Left: Spectral functions
 measured in quenched QCD simulations 
 on an anisotropic lattice with the spatial
  volume (1.25 fm)$^3$ for  the $J/\Psi$ channel 
  \cite{AH04}.
  Right: Spectral functions
 measured in quenched QCD simulations 
 on an anisotropic lattice with the spatial
  volume (1.34 fm)$^3$ for
   the $\eta_c$ channel  
  \cite{Jakovac07}.}
\label{asakawa}
\end{center}
\end{figure}

 Applications of MEM to the ${\rm s}\bar{\rm s}$ mesons
  at finite $T$ \cite{AHN03}
  and to the charmoniums at finite $T$
  \cite{AH04,Datta,Umeda05,Jakovac07} within the quenched QCD have been carried out 
  and  it was shown that 
  the $J/\Psi$ and $\eta_c$ survive
 even above $T_c$   as distinct peaks.
 Shown in the left (right) panel of
 Fig.\ref{asakawa} is the dimensionless spectral function   for 
 $J/\Psi$ ($\eta_{\rm c}$)  below and above $T_c$
 on $32^3\times (96-32)$ ($24^3\times (160-34) $)
 anisotropic lattices
 with the spatial volume $V = (1.25 {\rm fm})^3 $ ($V=(1.34 {\rm fm})^3$) 
 taken from  Ref.\cite{AH04} (Ref.\cite{Jakovac07}).
 The figures indicate that the low-lying s-wave resonances still survive 
 even at 1.5 $T_c$, which is
  consistent with the result in Fig.\ref{umeda}.
 Note that this result is not an artifact of the 
  small spatial volume, which was carefully checked 
 in \cite{Iida:2006mv}. 
 
 There are several topics which are interesting to be
  studied within the quenched QCD,
   such as the spectral functions of the charmoniums
  moving inside the hot medium
   \cite{charm-mom} and the spectral functions of
 bottomoniums above $T_{\rm c}$ \cite{bottom}.

  \begin{figure}[t]
\begin{center}
\includegraphics[width=7cm]{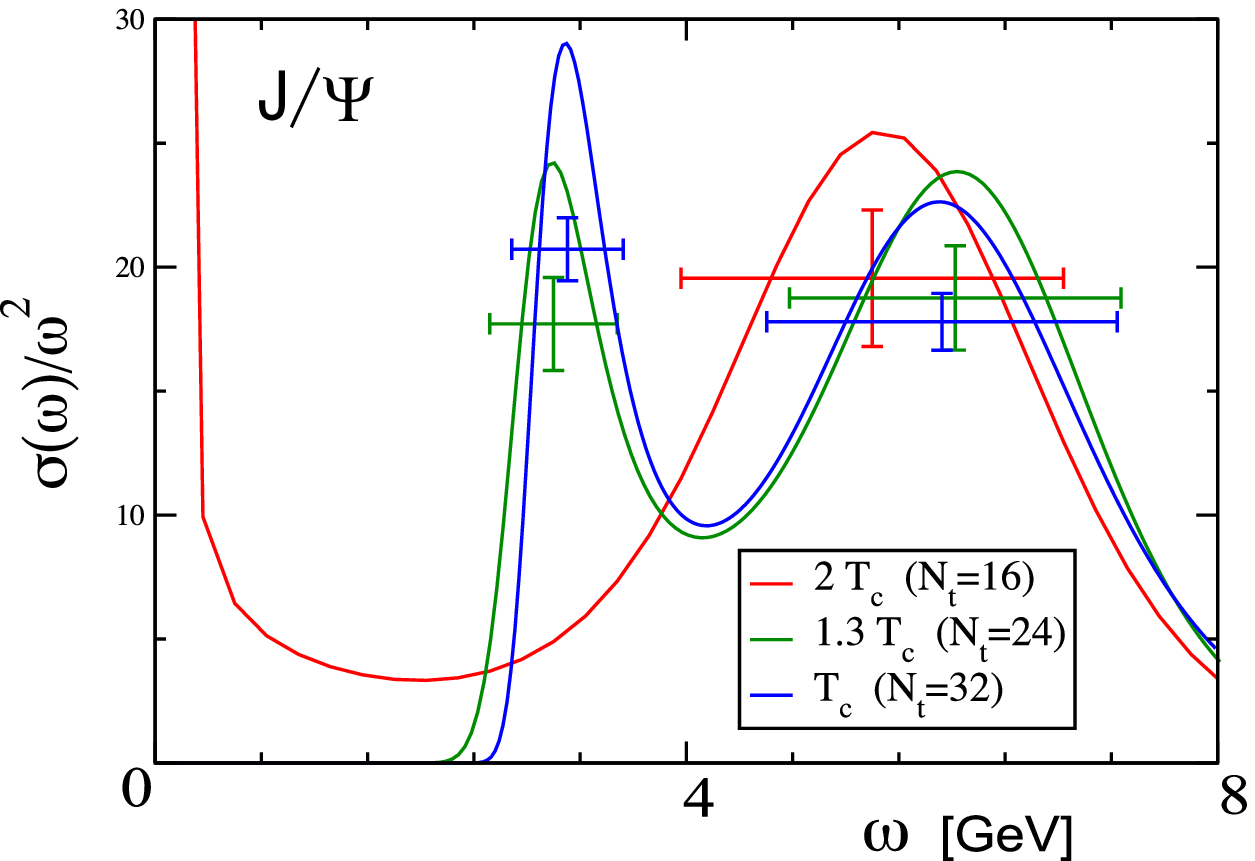}
\hspace{0.5cm}
\includegraphics[width=7cm]{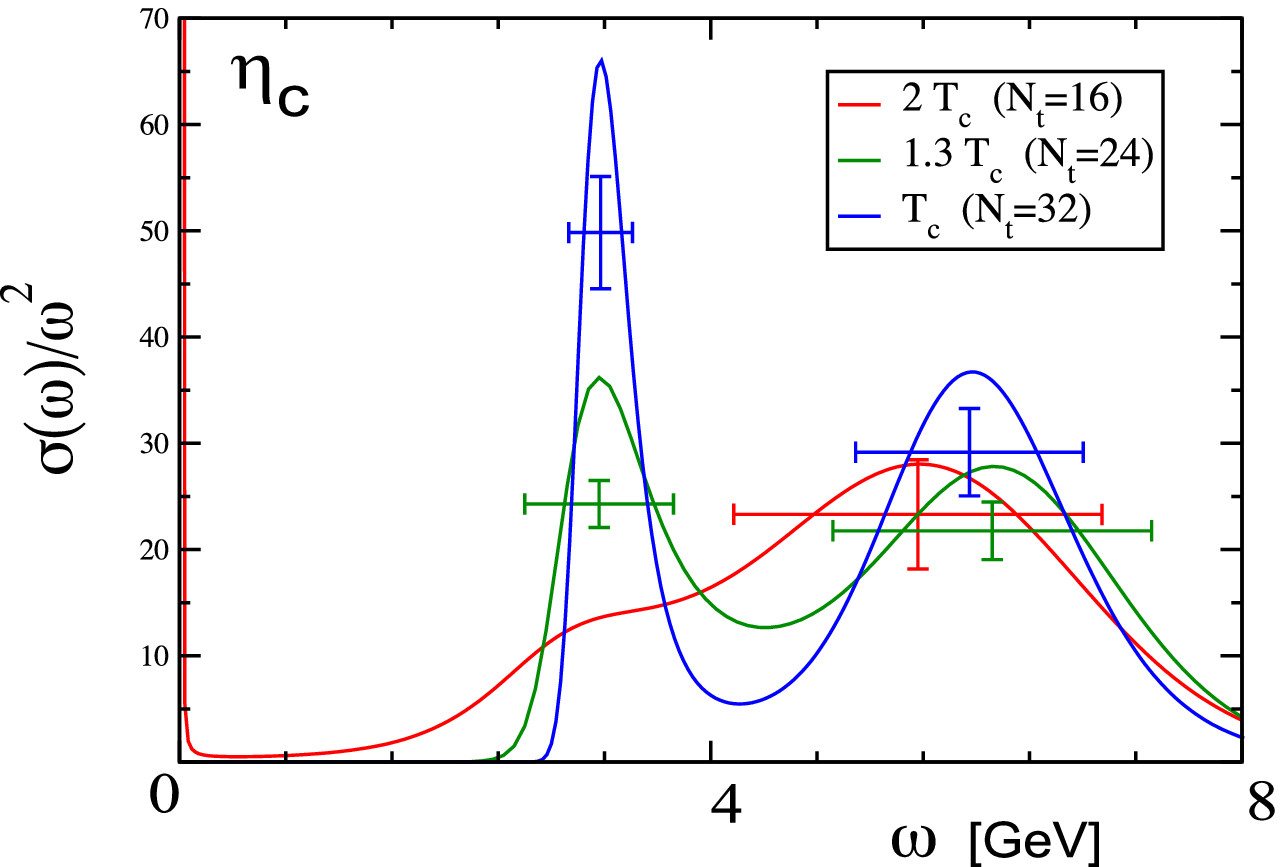}
\caption{Spectral functions in the vector and pseudo-scalar channels
 in 2-flavor QCD
 on $8^3 \times (32-16)$ anisotropic lattices.  
  The spatial lattice volume is  $V \sim$ (1.2 fm)$^3$ and the light quark mass
   corresponds to $m_{\pi}/m_{\rho}\sim 0.5$.
 The figures are adapted from  \cite{charm-full}. }
\label{aarts}
\end{center}
\end{figure}

\subsection{Charmonium spectral function in 2-flavor QCD}

 In quenched QCD, only gluons are thermally excited
 in the plasma, while, in full QCD, both quarks and gluons are thermally active.
 Because the number of degrees of freedom increase in full QCD, 
 the
  dissociation rate of the chamoniums will be larger.  On the other hand,
   the pseudo-critical temperature $T_{\rm pc}$ 
   in full QCD (160-200 MeV) is substantially 
   smaller than $T_{\rm c}\sim 270$ MeV in quenched QCD. Due to the 
  compensation of these effects, the ratio of the 
  dissociation temperature $T^*$ and the  (pseudo-)critical temperature
   in full QCD  is not much different from the ratio in quenched QCD 
   and could take a value of around 2 \cite{Hatsuda:2005nw}.  

  Aarts et al. has recently reported  an exploratory study of 2-flavor QCD
 on $8^3 \times (32-16)$ anisotropic lattices \cite{charm-full}.
 Shown in Fig.\ref{aarts} are the spectral functions of the 
  $J/\Psi$ and $\eta_{\rm c}$. 
   The results suggest that, even with the thermal excitations of
    light quarks, the low-lying charmoniums may  survive
    inside the quark-gluon plasma above $T_{\rm c}$ until they 
     dissociate away around $2 T_{\rm pc}$.

\section{Summary}
 
  We reviewed recent developments in lattice simulations
   on the equation of state, crossover nature of the
    thermal phase transition and
    the determination of the pseudo-critical temperature in (2+1)-flavor
   QCD.  Recent lattice
    results on the spectral properties of the heavy quarkoniums inside the 
    quark-gluon plasma are also summarized.  

\paragraph{{\bf Acknowledgements:}}
This work was partially  supported by Japanese MEXT Grant 
No. 18540253. The author thanks T. Umeda, S. Ejiri, F. Karsch and Z. Fodor for 
 useful information and discussions.


\newpage 

\section*{References}
\begin{thebibliography}{10}
\vspace{-0.3cm}
\bibitem{QGP}
Yagi K, Hatsuda T and Miake Y 2005 {\it  Quark-Gluon Plasma},
 (Cambridge Univ. Press, Cambridge).

\bibitem{asakawa-yazaki}
 Asakawa M and Yazaki K 1989 {\it Nucl.\ Phys.} {\bf A504} 668.\\
 Barducci A {\it et al.} 1989 {\it Phys.\ Lett.} {\bf B231} 463.
 
\bibitem{yamamoto}
Hatsuda T, Tachibana M, Yamamoto N, and Baym G 2006
 {\it Phys.\ Rev.\ Lett.}  {\bf 97} 122001.

\bibitem{schmidt06}
Schmidt C 2006  {\it PoS} {\bf LAT2006} 021 [hep-lat/0610116].

\bibitem{fermion-alg}
 Kennedy A D 2006 {\it arXiv:} hep-lat/0607038. 
Del Debbio L {\it et al.} 2006 {\it arXiv:} hep-lat/0610059.

\bibitem{clark06}
  Clark M A  2006 {\it arXiv:} hep-lat/0610048.

\bibitem{sakai06}
  Sakai S and Nakamura A  2006
  {\it PoS} {\bf LAT2005} 186  [hep-lat/0510100].

\bibitem{hirano06}
Hirano T and Gyulassy M 2006
 {\it Nucl.\ Phys.\ } {\bf A769} 71.

\bibitem{milc_eos_06}
  Bernard C {\it et al.} 2006 {\it arXiv:} hep-lat/0611031.

\bibitem{karsch_eos_06}
 Karsch F 2006  
 {\it J.\ Phys.\ Conf.\ Ser.\ } {\bf 46} 122.

\bibitem{aoki_eos_06}
 Aoki Y, Fodor Z, Katz S D and Szabo K K 2006
 {\it JHEP} {\bf 0601} 089.
  
 \bibitem{aoki_sus_06}
Aoki Y, Endrodi G, Fodor Z, Katz S D, and Szabo K K 2006
 {\it Nature} {\bf 443} 675.  


\bibitem{rebhan04}
  Kraemmer U and Rebhan A 2004 {\it Rept.\ Prog.\ Phys.\  } {\bf 67} 351.

\bibitem{MILC06_tc}
Bernard C {\it et al.} 2005 {\it Phys. Rev.} {\bf D71} 034504.

 
\bibitem{RBCB06_tc}
 Cheng M {\it et al.} 2006 {\it Phys.\ Rev.\ }{\bf D74} 054507.
  
\bibitem{WB06_tc} 
Aoki Y, Fodor Z, Katz S D, and Szabo K K 2006
{\it Phys.\ Lett.\ } {\bf B643} 46.

\bibitem{ukawa93}
Ukawa A 1993 {\it Lectures on Lattice QCD at Finite Temperature},
 (Uehling Summer School, Seattle).

\bibitem{maezawa06}
Maezawa Y {\it et al.}
 2007 {\it arXiv:} hep-lat/0702005.

\bibitem{HPQCD-UKQCD}
Gray A {\it et al.} (HPQCD and UKQCD Coll.) 2005
{\it Phys. Rev.} {\bf D72} 094507.

\bibitem{CP-PACS-JLQCD}
Ishikawa T {\it et al.} (CP-PACS and JLQCD Coll.) 2006
 {\it PoS} {\bf LAT2006} 181 [hep-lat/0610050].


\bibitem{MS}
  Matsui T and Satz H 1986
 {\it  Phys.\ Lett.\ } {\bf B178} 416.\\
  Hashimoto T, Hirose K, Kanki T and Miyamura O 
  1986 {\it Phys.\ Rev.\ Lett.\ } {\bf 57} 2123.

\bibitem{umeda}
 Umeda T, Katayama R, Miyamura O and Matsufuru H 2001
 {\it Int.\ J.\ Mod.\ Phys.\ }  {\bf A16} 2215.

\bibitem{HK85}
Hatsuda T and Kunihiro T 1985 {\it Phys.\ Rev.\ Lett.\ } {\bf 55} 158.
DeTar C 1985 {\it Phys.\ Rev.\ } {\bf D32} 276.

\bibitem{AHN01}
 Asakawa M, Hatsuda T and Nakahara Y 2001
 {\it Prog.\ Part.\ Nucl.\ Phys.\ } {\bf 46} 459.
 
\bibitem{AHN03}
 Asakawa M, Hatsuda T and Nakahara Y 2003
 {\it Nucl. Phys. Proc. Suppl. } {\bf 119} 481.

\bibitem{AH04}
Asakawa M and Hatsuda T 2004
 {\it Phys. Rev. Lett. } {\bf 92} 012001;
 {\it J. Phys. } {\bf G30} S1337.

\bibitem{Datta}
 Datta S, Karsch F, Petreczky P and Wetzorke I  2004
 {\it  Phys. Rev. } {\bf D69} 094507.
 
\bibitem{Umeda05}
 Umeda T, Nomura K and  Matsufuru H 2005
 {\it  Eur. Phys. J.} {\bf C39S1} 9.

\bibitem{Jakovac07}
  Jakovac A, Petreczky P,  Petrov K and Velytsky A 2007
 {\it Phys.\ Rev.\ } {\bf D75} 014506.

\bibitem{Iida:2006mv}
 Iida H, Doi T, Ishii N, Suganuma H and Tsumura K 2006 
 {\it Phys.\ Rev.\  }  {\bf D74} 074502.
 
\bibitem{charm-mom}
  Datta S, Karsch F, Wissel S, Petreczky P and Wetzorke I 2004
 {\it arXiv:} hep-lat/0409147.\\
 Aarts G, Allton C, Foley J, Hands S and Kim S 2006
  {\it arXiv:} hep-lat/0610061.
 
\bibitem{bottom}
  Datta S, Jakovac A, Karsch F  and Petreczky P 2006
   {\it AIP Conf.Proc.} {\bf 842} 35 [hep-lat/0603002].

\bibitem{Hatsuda:2005nw}
  Hatsuda T T 2006 {\it Int.\ J.\ Mod.\ Phys.\ } {\bf A21} 688
   [hep-ph/0509306].

\bibitem{charm-full}
  Aarts G {\it et al.} 2006
  {\it PoS} {\bf LAT2006} 126 [hep-lat/0610065].
 

\end{thebibliography}
\end{document}